\documentclass[12pt,preprint]{aastex}

\def\a0{a_0}
\def\msun{M_{\odot}}

\def\deg{^o}
\def\kpc{{~\rm kpc~}}

\def\mol{$M/L~$}

\def\cmss{cm s$^{-2}$}
\def\deg{^o}

\begin{document}

\title{{\bf MOND and the mass discrepancies in tidal dwarf galaxies}}
\author{Mordehai Milgrom }
\affil{ Center for Astrophysics, Weizmann Institute}

\begin{abstract}
I consider in light of MOND the three debris galaxies discussed
recently by Bournaud et al.. These exhibit mass discrepancies of a
factor of a few within several scale lengths of the visible galaxy,
which, arguably, flies in the face of the cold dark matter paradigm.
I show here that the rotational velocities predicted by MOND agree
well with the observed velocities for each of the three galaxies,
with only the observed baryonic matter as the source of gravity.
There is thus no need to invoke a new form of baryonic,
yet-undetected matter that dominates the disc of spiral galaxies, as
advocated by Bournaud et al.. I argue on other grounds that the
presence of such ubiquitous disc dark matter, in addition to cold
dark matter, is not likely.
\end{abstract}


\section{Introduction}
Bournaud et al. (2007) have recently reported on three tidal dwarf
galaxies that apparently formed in the debris of the collision of
two galaxies: possibly of NGC 5291 with another galaxy. For reasons
explained cogently by the authors hardly any cold dark matter from
the parent galaxies, if it existed there in the first place, is
expected to be found in these debris. If this is indeed so, the cold
dark matter paradigm (CDM) predicts no mass discrepancy in these
dwarfs, contrary to what has been reported by Bournaud et al.
(2007): they find in all three dynamical masses within several scale
lengths that exceed the the observed baryonic masses by a factor of
a few.
\par
The dark-matter (DM) paradigm and MOND (Milgrom 1983a,b, see Sanders
\& McGaugh 2002, and Bekenstein 2006 for reviews) differ greatly as
regards the origin and nature of mass discrepancies they predict in
galaxies. In MOND, these discrepancies in a given galaxy are
predicted exactly from the presently observed mass distribution, and
are oblivious to the exact formation process of the galaxy or the
ensuing history. They are predicted, and are observed, to follow
some well defined, strict regularities: galactic analogs of Kepler's
laws, such as the relation between total baryonic mass and the
asymptotic rotation speed (the baryonic Tully Fisher relation), or
the onset of the discrepancy at a fixed value of the centrifugal
acceleration (see e.g., Milgrom 2002 for a discussion of these
predictions, and McGaugh 2006 for a discussion of observational
tests). In contrast, in the DM paradigm the mass discrepancies are
ratios of total (dark matter plus baryons) to the baryonic mass, and
depend strongly on the particular history of the galaxy since the DM
and the baryons are subject to different influences. The formation
process itself, subsequent cannibalism, mergers, and ejection of
baryons by cataclysmic events, such as supernovae, all greatly
affect the resulting mass discrepancies. This is why I believe that
the CDM paradigm is inherently incapable of ever predicting rotation
curves of individual galaxies in the way that MOND does: for most
galaxies we simply cannot know the crucial elements of evolution a
given galaxy underwent.
\par
The three reported dwarfs, and possibly others like them, are an
exception: if they indeed formed as described by Bournaud et al.
then their mass discrepancies in CDM can be predicted with some
certainty because the collision that led to their formation erased
the imprints of earlier history. According to the simulations of
Bournaud et al., whatever the DM halo of the parent galaxies was
like, as long as it was spheroidal--as predicted by CDM--the debris
galaxies would have hardly any DM in them, and should exhibit
practically no mass discrepancies. This is also in line with earlier
simulations referenced in the online appendices of Bournaud et al.,
and is easy to understand qualitatively. In contrast, all three
dwarfs are predicted by MOND to show appreciable mass discrepancies
since they are measured to have low accelerations that are rather
deep in the MOND regime. I show below that the MOND predictions are
indeed born out by the observations of Bournaud et al.. Gentile et
al. (2007a) have recently performed a more detailed analysis, and
reach the same conclusions as regards the performance of MOND.
\par
Bournaud et al. consider it a more likely explanation of the mass
discrepancy in the dwarfs, that they actually do contain large
quantities of yet-undetected matter. They advocate that at least one
of the galaxies partaking in the collision that begot the debris had
large quantities of DM in their discs--with a mass typically a few
times that of the visible baryons. Since this DM cannot be the
putative cold dark matter, which form spheroidal halos and does not
settle into galactic discs, Bournaud et al. opt for cold molecular
hydrogen, which has been considered earlier as the DM in galaxies
and clusters (e.g., Pfenniger Combes \& Martinet 1994).

In section 2 I describe the MOND results and compare them with the
observations. In section 3 I discuss the results and contest the
hypothesis of large quantities of disc dark matter in galaxies.

\section{MOND rotation curves}

To calculate the predicted MOND rotation velocities, $V$, I use the
Newtonian velocities, $V_N$, read from Fig. 1 of Bournaud et al.
(2007) in the MOND relation (Milgrom 1983b)

$$\mu(V^2/r\a0)V^2/r=V^2_N/r. $$

Here $\mu(x)$ is the extrapolating function, which I take here to be
$\mu(x)=x(1+x^2)^{-1/2}$  and I take the acceleration constant of
MOND to have the value $\a0=1\times 10^{-8}$\cmss (e.g. Bottema et
al. 2002). Choosing another form of $\mu(x)$ will affect the
predictions for the larger radii only a little since the
accelerations there are rather smaller than $\a0$ where $\mu$ has to
be nearly linear for all forms. The velocities at the inner radii
will be affected, but as explained below these anyway depend
crucially on the model adopted for the baryon mass distribution.

Figures 1-3 show for each galaxy the Newtonian velocities calculated
from the distribution of the visible matter alone as modeled by
Bournaud et al., the MONDian speeds calculated from the above
equation for the same mass distribution, and the observed rotational
speeds. The latter are the average of the approaching and receding
velocities as given in Bournaud et al.. (The differences between the
two sides are much smaller than the errors.) For clarity's sake I
have not marked the error bars for the MONDian speeds since they
anyhow all fall within the error bars of the measured values.
\par
In the above procedure I have adopted all the system parameters (the
distance,  the inclinations of the galaxies, the assumed \mol values
for the stellar contributions, etc.) as taken by Bournaud et al.,
and no attempt was made to improve the agreement by best fitting for
these. The baryonic mass of these galaxies are dominated by gas so
the exact \mol value is rather immaterial. I have also not corrected
for asymmetric drift (i.e., those due to velocity dispersions),
which is expected to be rather small for these galaxies (certainly
much smaller than the quoted errors). Similarly, I have ignored
possible corrections due to the external-field-effect (EFE) in MOND
(Milgrom 1983a, Brada \& Milgrom 2000a,b, Angus \& McGaugh 2007, and
Wu et al. 2007 ). I estimate that it is rather unimportant here
(i.e., the acceleration field of NGC 5291 itself, and of other
masses, at the dwarfs positions is smaller than the internal
accelerations in the dwarfs themselves). Gentile et al. (2007a)
study this issue in more detail. They find that indeed the impact of
the EFE is marginal inside the last measured point. My own estimate
of the EFE due to NGC 5291 is even smaller than theirs. Gentile et
al. calculate the far field of NGC 5291 by assuming that the
(deprojected) HI line width ($\Delta V_{20}$) given in Malphrus et
al. (1997) represents twice the asymptotic rotational speed of that
galaxy. This corresponds to a baryonic mass of
$3.2\times10^{11}\msun$, which, the galaxy being rather devoid of
gas, corresponds to $M/L_B\approx 16 (M/L_B)_{\odot}$; this is much
too high ($\Delta V_{20}/2$ could easily overestimate the asymptotic
rotational speed for various reasons; for example, since the galaxy
is rather Newtonian in the inner parts, the maximum rotation speed
can be quite higher than the asymptotic one). I started from the
luminosity of $L_B\approx 2\times 10^{10}L_\odot$ and assumed
$M/L_B\approx 5 (M/L_B)_{\odot}$, appropriate for the color of the
galaxy ($B-V\approx 1$). The same mass is gotten with $M/L_K= 1
(M/L_K)_{\odot}$, and it corresponds to a galaxy mass 3.2 times
smaller than that effectively used by Gentile et al.. Also, whereas
they used the projected distances from NGC 5291 to the dwarfs,
taking a mean of $65\kpc$ as the actual distances, I used 3-D
distances using the position of the galaxy with respect to the ring
from the model of Bournaud et al.: $114\kpc$ for N, $118\kpc$ for S,
and $140 \kpc$ for SW (F. Bournaud 2007, private communication). All
together my estimates of the field of the galaxy at the position of
the dwarfs are 3.2--3.8 times smaller than that of Gentile et al.,
rendering the effect rather negligible within the presently observed
dwarfs.

Bournaud et al. argue convincingly for inclinations around
$i=45\deg$ for all three dwarfs stating that they should be almost
aligned with the ring they are embedded in.  A large part of the
indicated errors on the observed velocities reflect the uncertainty
in $i$. This contribution to the errors should than be viewed as an
uncertainty in the normalization of the velocity curve, not as
errors on individual points.  I show in Figure 1 that for one case
(NGC 5291N), increasing the inclination to $55\deg$ indeed improves
greatly the agreement bringing the MOND velocities into practical
coincidence with the observed ones.
\par
Gentile et al. (2007a) suggest that the inclinations of the dwarfs
may differ from the model inclinations of $45\deg$ because the EFE
induces some precession of the disc. They thus also performed MOND
fits with the inclinations left free, and indeed the best fit values
differ somewhat from $45\deg$. This is also shown in my Figure 1
demonstrating that the best value for NGC 5291N is nearer $55\deg$.
This point might require further checking; but, as explained above,
my arguably more realistic estimate of the external field effect is
rather smaller than theirs and would give an estimate of the
precession period that is much longer than the time since the dwarfs
formed, so I do not expect this to be an important effect.

\begin{center}
\begin{figure}
\begin{tabular}{rl}
\tabularnewline
\includegraphics[width=0.5\columnwidth]{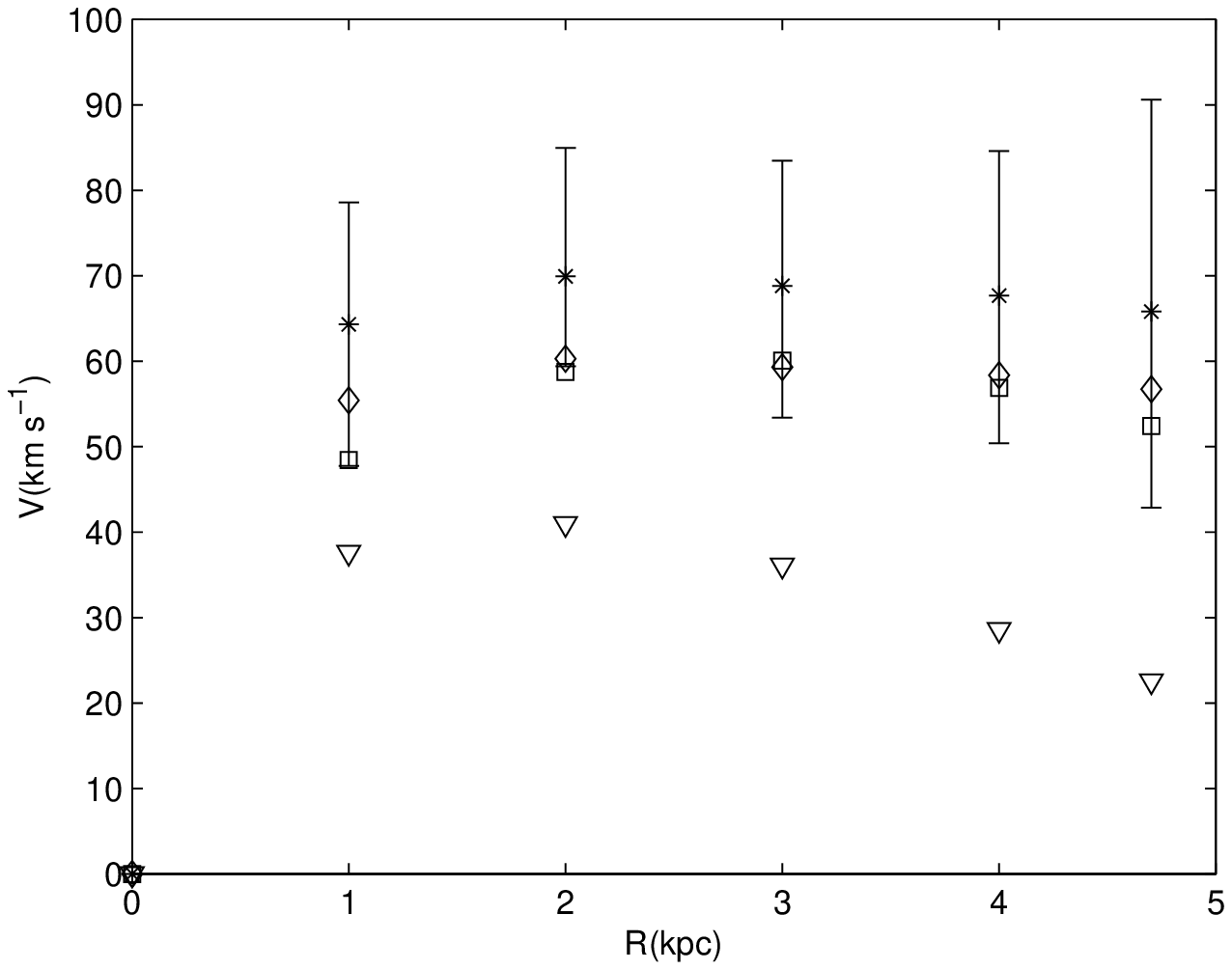} &
\includegraphics[width=0.5\columnwidth]{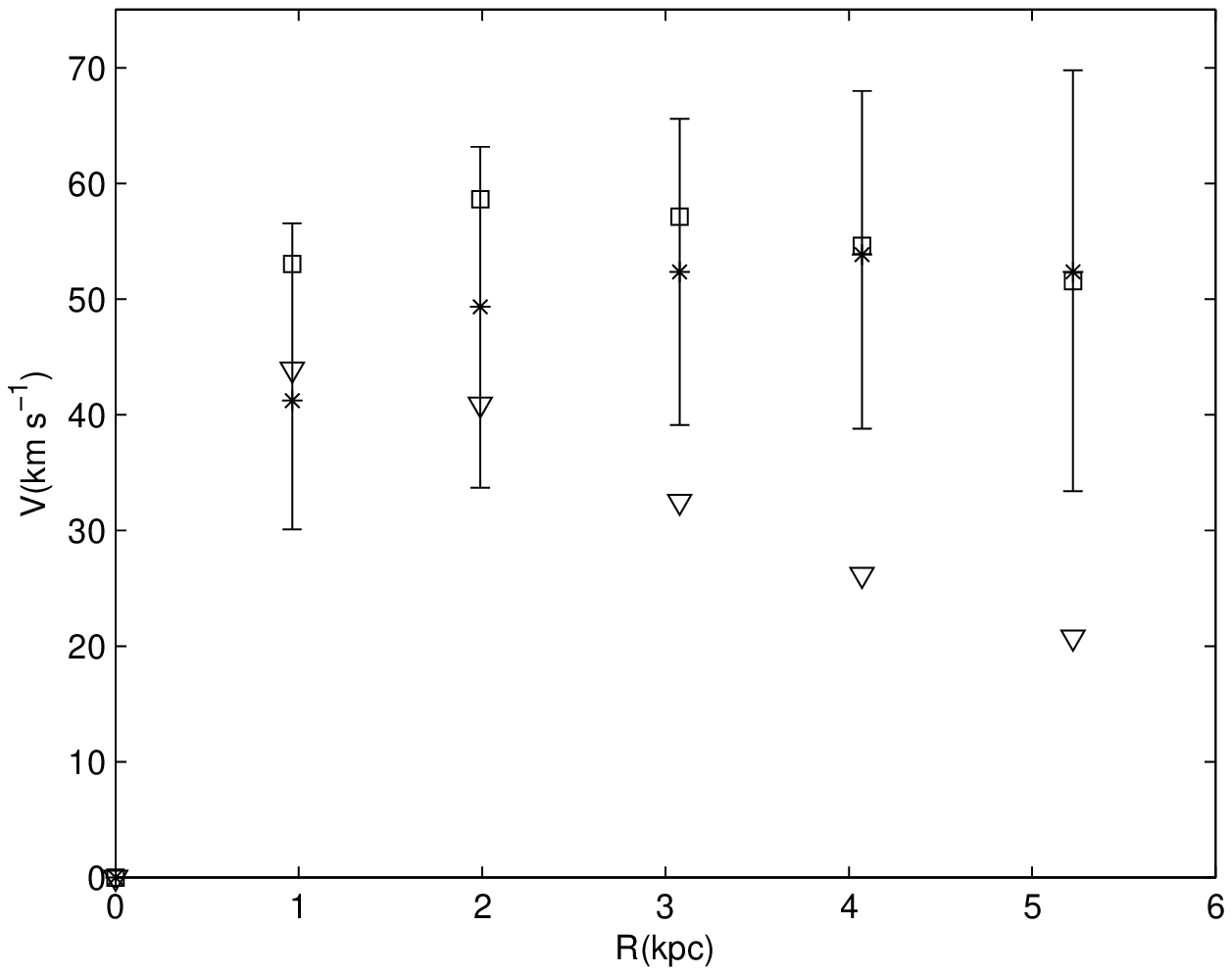} \\
\includegraphics[width=0.5\columnwidth]{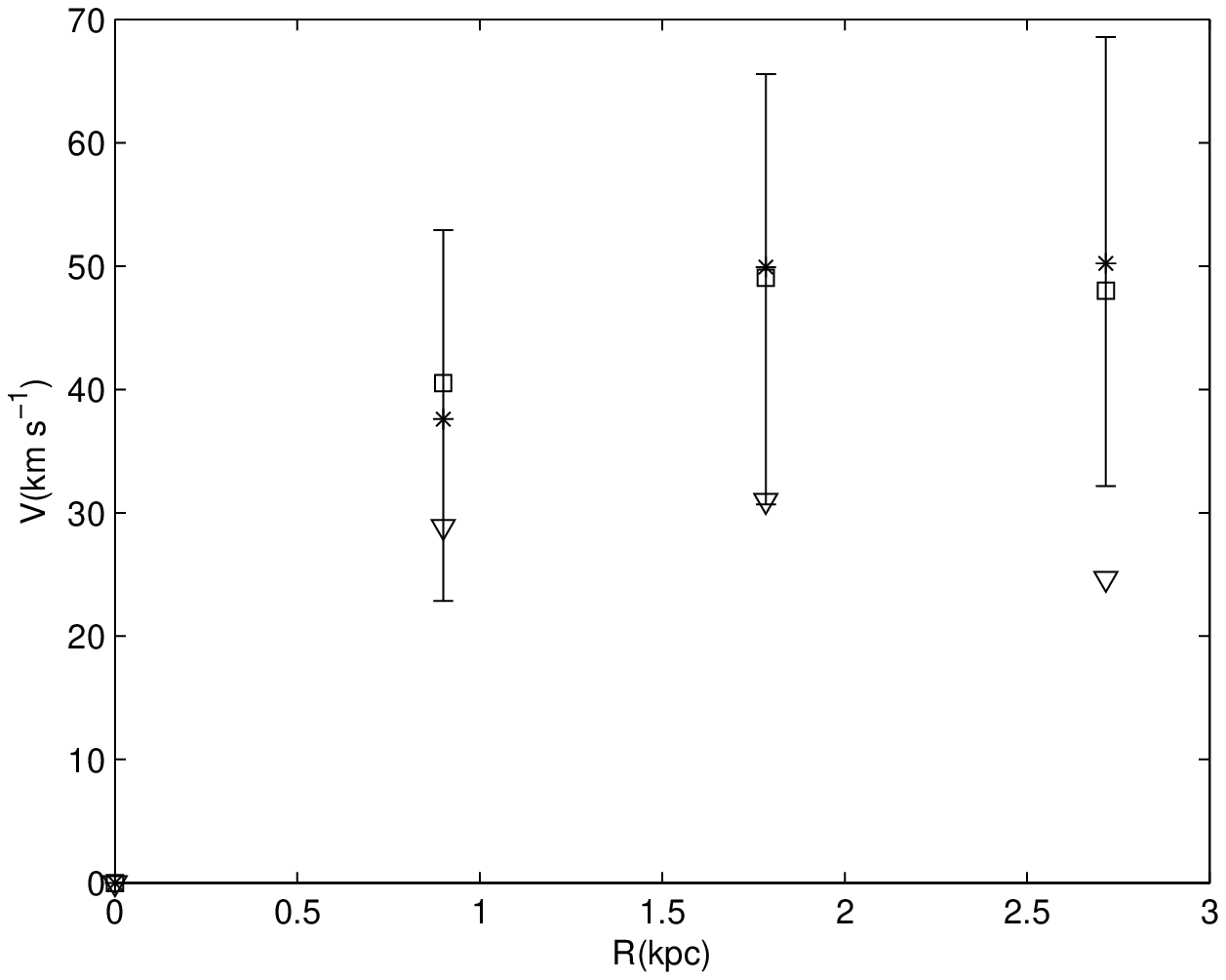}\\
\end{tabular}\par
\caption{The rotation curves for the three dwarfs: NGC 5291N (upper
left), and 5291S (upper right), 5291SW (lower left). The measured
velocities, assuming the nominal inclination of $45\deg$, are marked
by stars and are shown with their error bars. The calculated
Newtonian velocities are marked by inverted triangles. The predicted
MOND velocities are marked by squares. The error bars for the last
two are omitted for clarity's sake. I also show, for NGC 5291N only,
as diamonds, the measured velocities for an assumed inclination of
$i=55\deg$.}
\end{figure}
\end{center}

\section{Discussion}
We see that the MOND predictions are in very good agreement with
the measured speeds. It should be noted that the baryonic
Newtonian curves given in Bournaud et al. are not based on an
actual measurement of the observed baryonic distribution. It is
based on taking the total amount of observed baryon and assuming a
model for their spatial distribution. The exact values of
Newtonian velocities, and hence the MOND values I deduce from them
for the inner radii depend on the exact distribution assumed.
However, at larger radii, which are already quite beyond the
baryonic mass concentration, the velocity values are rather
independent of the assumed distribution of baryons. And, after
all, this is where the impact of the comparison with both CDM and
MOND predictions comes from.
\par
The uncertainties in the analysis of the individual dwarfs are
relatively large. This can be traced mainly to the fact that they
are much farther than the dwarf spirals of similar properties that
have been analyzed before (e.g.,  de Blok \& McGaugh 1998, Gentile
et al. 2007b, Milgrom and Sanders 2007). However, while
individually, the data is not up to the highest standards, all
three dwarfs speak in the same voice: all showing a discrepancy of
similar magnitude developing at larger radii, with the observed
velocities flattening off in just the way and magnitude predicted
by MOND. Thus, the collective conclusion is stronger than the
individual ones separately. In addition, the particular importance
of these systems lies in their unique potential for
differentiating between CDM and MOND.

\par
Bournaud et al. propose another explanation of the mass discrepancy
in the three dwarfs: one of the colliding galaxies, which
contributed the gas to the debris, also harbored large quantities of
some form of DM (not CDM) in its disc, several times more massive
than the visible baryons. This DM could then have found its way into
the tidal dwarfs, giving rise to the observed mass discrepancy. The
candidate they advocate is cold, difficult to detect H$_2$. Unless
we want to assume that the presence of such molecular DM in the disc
of the parent galaxy is a rare occurrance, this DM has to be
ubiquitous in disc galaxies, as indeed is proposed by Bournauad et
al.. They note, however, that this would require a very large
conversion factor from the observed CO to H$_2$: an order of
magnitude larger than what is known for galaxies in general (see
also the caveats listed by Elmergreen 2007).
\par
I feel that the presence of large quantities of a new component of
DM in the disc of spiral galaxies, in addition to the dominant CDM,
is unlikely on additional grounds. It is known that the mass and the
distribution of the DM in disc galaxies are strongly correlated with
those of observed baryons through the relations predicted by MOND.
For example, the total mass of the latter, $M_{vis}$, is strongly
correlated with the asymptotic rotational speed, $V_{\infty}$ (which
is determined mostly by the DM) via the MOND relation (Milgrom
1983b)
$$V^4_{\infty}=\a0 GM_{vis}.$$ This MOND relation (aka the baryonic
Tully-fisher relation) is found to hold over some 5 orders of
magnitude in galactic mass (McGaugh 2005,2006). Another correlation
is the onset of the mass discrepancy (equality of the contributions
of visible and dark matter) at a fixed value, $\a0$, of the
centrifugal acceleration (see McGaugh 2006 for a recent test of
this). But the mother of them all is the fact that the visible
matter distribution determines the full rotation curve of a galaxy,
which in the dark matter paradigm is determined by both components
and is dominated by DM in the outer parts. These are all
observational facts whatever the interpretation of MOND is. In the
context of CDM these correlations require various independent
conspiracies between the baryons and the CDM, conspiracies whose
origin remains a mystery (see Milgrom 2002 for a more extensive list
of these conspiracies and an explanation of why they are
independent).
\par
As emphasized many times before, these predicted MOND relations as
traditionally formulated are exactly valid only for completely
isolated systems. In the presence of an external field these are
modified in a manner that is also predicted by MOND (provided the
external field is known). However, for many systems the external
field effect enters importantly only at rather large radii and
leaves a large range of radii for which these predictions are valid
with high accuracy (for the Milkey Way it is expected to be
important only beyond a few hundred kiloparsecs). The effects of
external fields are deemed quite unimportant, as far as I know, for
all rotation curve analyses published to date, and hence are also
unimportant, for example, in the results of McGaugh 2006 quoted
above. Some exceptions concern galaxies in the cores of galaxy
clusters, where the external (cluster) fields are of order the of
$\a0$, and low acceleration systems (such as dwarf spheroidals or
diffuse globular clusters) in the field of galaxies.

\par
If one now adds another epicycle to the DM paradigm in the form of a
dominant, baryonic disc component, the observed correlations, by
which only the sub-dominant, visible-baryons component determines
everything, would require an even more involved, three-headed
conspiracy. This is not an argument that definitely excludes the
molecular-DM-plus-CDM hypothesis, but it does diminish its
likelihood.
 \par
  The above argument poses a difficulty for the double
DM hypothesis even before we consider the tidal dwarfs themselves.
In addition, the present rotation curve results show that the three
dwarfs also satisfy the above correlations. This adds another
dimension to the above argument: if we accept that the dwarfs formed
in a very different way from that of most other galaxies, and that
their matter component mixture is very different, why should they
still satisfy the same relations? In the dwarfs' case these would be
relations between the visible baryons and the molecular DM, while in
general they would be relations between the visible baryons and the
combined molecular-plus-cold DM. Of course, this latter part of the
argument assumes the robustness of the results and interpretation of
Bournaud et al., including their deduced inclinations. It is also
based, at the moment, on only the three galaxies discussed here. The
argument would clearly benefit from further substantiation, and the
examination of more tidal dwarfs. Until then it remains a tentative
difficulty for the double DM hypothesis.

\acknowledgements I am grateful to Rainer Plaga for pointing out to
me the potential in the results of Bournaud et al. to discriminate
between MOND and CDM. I also appreciate comments from Frederic
Bournaud and from the referee. The research was supported by a
center of excellence grant from the Israel Science Foundation.


\begin{thebibliography}{}

\bibitem[Agnus \& McGaugh 2007]{agnus07}  Angus, G.W. \& McGaugh, S.S. 2007, arXiv:0704.0381
\bibitem[Bekenstein 2006]{bek06} Bekenstein, J.D. 2006, Contemporary Physics 47, 387
\bibitem[Bottema et al. 2002]{bottema02}Bottema, R., Pesta\~{n}a, J. L. G., Rothberg, B., Sanders, R. H. 2002, AA, 393, 453
\bibitem[Bournaud et al. 2007]{bournaud} Bournaud, F., Duc, P.-A.,
Brinks,E.,  Boquien, M., Amram, M., Lisenfeld, U., Koribalski, B.S.,
Walter, F., Charmandaris, V. 2007, Science, 316, 1166
\bibitem[Brada \& Milgrom 2000a]{bm00a}Brada, R. \& Milgrom, M. 2000a ApJ, 541, 556
\bibitem[Brada \& Milgrom 2000b]{bm00b}Brada, R. \& Milgrom, M. 2000b ApJ, 531, L21
\bibitem[de Blok \& McGaugh 1998]{dbmg98}de Blok, E. \& McGaugh, S.S. 1998, ApJ, 508, 132
\bibitem[Elmergreen 2007]{elmer07}Elmergreen, B.G. 2007, Science (perspectives), 316,
1132
\bibitem[Gentile et al. 2007a]{gentile07a} Gentile, G., Famaey, B.,
Combes, F., Kroupa, P., Zhao, H.S., \& Tiret, O. 2007a, AA, in
press, arXiv:0706.1976
\bibitem[Gentile et al. 2007b]{gentile07b} Gentile, G., Salucci, P.,  Klein, U., and
Granato, G. L. 2007b MNRAS 375, 199
\bibitem[Malphrus et al. 1997]{malphrus97} Malphrus, B.K., Simpson,
C.E., Gotesman, S.T., \& Hawarden, T.G. 1997, AJ 114, 1427
\bibitem[McGaugh 2005]{mcg05}  McGaugh, S.S. 2005, ApJ, 632, 859
\bibitem[McGaugh 2006]{mcg06}  McGaugh, S.S. 2006,
astro-ph/0606351
\bibitem[Milgrom 1983a]{mil83a} Milgrom, M. 1983a, ApJ, 270, 365
\bibitem[Milgrom 1983b]{mil83b} Milgrom, M. 1983b, ApJ, 270, 371
\bibitem[Milgrom 2002]{mil02} Milgrom, M. 2002, New Astron.Rev. 46 741
\bibitem[Milgrom \& Sanders 2007]{ms07}Milgrom, M. \& Sanders, R.H. 2007 ApJ, 658, L17
\bibitem[Pfenniger Combes \& Martinet 1994]{pfenniger94}Pfenniger,
D., Combes, F., \& Martinet, L. 1994, AA, 285, 79
\bibitem[Sanders \& McGaugh 2002]{sm02}Sanders, R.H. \& McGaugh, S.S. 2002, ARA\&A, 40, 263
\bibitem[Wu et al. 2007]{wu07} Wu, X., Zhao, H.S., Famaey, B.,
Gentile, G., Tiret, O., Combes, F., Angus, G.W., \& Robin, A.C.
2007, arXiv:0706.3703v2



\end{thebibliography}
\end{document}